\documentclass[12pt,thmsa,notitlepage]{article}
\usepackage{amsmath}
\usepackage{amssymb}
\usepackage{amsfonts}
\usepackage{sw20lart}
\usepackage[onehalfspacing]{setspace}

\input{tcilatex}

\begin{document}

\title{ \textbf{Coupling Constant Unification in Extensions of Standard Model%
}}
\author{{\large Ling-Fong Li}$,$ and {\large Feng Wu} \\
{\small Department of Physics, Carnegie Mellon University, Pittsburgh, PA
15213}}
\maketitle

\begin{abstract}
Unification of electromagnetic, weak, and strong coupling constants is
studied in the extension of standard model with additional fermions and
scalars. It is remarkable that this unification in the supersymmetric
extension of standard model yields a value of Weinberg angle which agrees
very well with experiments. We discuss the other possibilities which can
also give same result.
\end{abstract}

One of the attractive features of the Grand Unified Theory is the
convergence of the electromagnetic, weak and strong coupling constants at
high energies and the prediction of the Weinberg angle\cite{Gut},\cite{GQW}.
This lends a strong support to the supersymmetric extension of the Standard
Model. This is because the Standard Model without the supersymmetry, the
extrapolation of 3 coupling constants from the values measured at low
energies to unification scale do not intercept at a single point while in
the supersymmetric extension, the presence of additional particles, produces
the convergence of coupling constants elegantly\cite{SUSY}, or equivalently
the prediction of the Weinberg angle agrees with the experimental
measurement very well\cite{Exp}. This has become one of the cornerstone for
believing the supersymmetric Standard Model and the experimental search for
the supersymmetry will be one of the main focus in the next round of new
accelerators.

In this paper we will explore the general possibilities of getting coupling
constants unification by adding extra particles to the Standard Model\cite%
{early} to see how unique is the Supersymmetric Standard Model in this
respect\cite{Marciano}. We will compute the contribution of various types of
new particles to the evolution of the coupling constants and study their
effects on the unification. These results will also be useful for checking
models with extra particles to see whether they have a satisfactory
unification of coupling constants.

\section{ Evolution of Coupling Constants}

In this section we will set up the framework to discuss the unification in
the $SU\left( 5\right) $ type of model. Recall that the evolution of the
coupling constants $g_{3},g_{2},g_{1}$ of the subgroups $SU\left( 3\right)
_{C},$ $SU\left( 2\right) _{W},U\left( 1\right) _{Y}$ \ at one loop are
given by%
\begin{equation}
\dfrac{1}{g_{i}^{2}\left( \mu \right) }=\dfrac{1}{g_{i}^{2}\left( \mu
_{0}\right) }+2b_{i}\ln \left( \dfrac{\mu }{\mu _{0}}\right) ,\qquad i=1,2,3
\end{equation}%
where $b_{i}$ are the coefficients in $\beta -$function of the
renomalization group equations\cite{RGE},%
\begin{equation}
\dfrac{dg_{i}}{d\left( \ln \mu \right) }=-b_{i}g_{i}^{3},
\end{equation}%
Note that the coupling constants $g_{2},g_{1}$ are related to the usual
coupling constants, $g,g^{\prime }$ in $SU\left( 2\right) \times U\left(
1\right) $ by%
\begin{equation}
g=g_{2},\qquad g^{\prime }=\sqrt{\dfrac{3}{5}}g_{1}
\end{equation}%
and Weinberg angle is given by%
\begin{equation}
\sin ^{2}\theta _{W}=\dfrac{g^{\prime 2}}{g^{2}+g^{\prime 2}}
\end{equation}%
At the unification scale $M_{X},$ we have 
\begin{equation}
g_{1}\left( M_{X}\right) =g_{2}\left( M_{X}\right) =g_{3}\left( M_{X}\right)
=g_{5}
\end{equation}%
where $g_{5}$ is the gauge coupling of the $SU\left( 5\right) $ group. Using
the relation, $e=g\sin \theta _{W},$ we can write%
\begin{equation}
\dfrac{1}{\alpha _{3}\left( \mu \right) }=\dfrac{1}{\alpha _{5}}+8\pi
b_{3}\ln \left( \dfrac{\mu }{M_{X}}\right)  \label{unified1}
\end{equation}%
\begin{equation}
\dfrac{1}{\alpha \left( \mu \right) }\sin ^{2}\theta _{W}=\dfrac{1}{\alpha
_{5}}+8\pi b_{2}\ln \left( \dfrac{\mu }{M_{X}}\right)  \label{unified2}
\end{equation}%
\begin{equation}
\dfrac{3}{5}\dfrac{1}{\alpha \left( \mu \right) }\cos ^{2}\theta _{W}=\dfrac{%
1}{\alpha _{5}}+8\pi b_{1}\ln \left( \dfrac{\mu }{M_{X}}\right)
\label{unified3}
\end{equation}%
where $\alpha =\dfrac{e^{2}}{4\pi },$ $\alpha _{3}=\dfrac{g_{3}^{2}}{4\pi },$
$\alpha _{5}=\dfrac{g_{5}^{2}}{4\pi }.$ Eliminating the unification scale $%
M_{X}$ from Eqs(\ref{unified1},\ref{unified3}) we get%
\begin{equation}
\sin ^{2}\theta _{W}=\dfrac{1}{\left( 1+\dfrac{8}{5}\overset{\_}{b}\right) }%
\left[ \dfrac{3}{5}\overset{\_}{b}+\left( \dfrac{\alpha }{\alpha _{3}}%
\right) \right]
\end{equation}%
where%
\begin{equation}
\overset{\_}{b}=\dfrac{b_{3}-b_{2}}{b_{2}-b_{1}}  \label{b-bar1}
\end{equation}%
and all coupling constants $\alpha _{i}$ are evaluated at $\mu =M_{Z}.$ For
convenience, we can solve $\overset{\_}{b}$ in terms of $\sin ^{2}\theta
_{W},$%
\begin{equation}
\overset{\_}{b}=\dfrac{\left( \dfrac{\alpha }{\alpha _{3}}-\sin ^{2}\theta
_{W}\right) }{\left( \dfrac{8}{5}\sin ^{2}\theta -\dfrac{3}{5}\right) }
\end{equation}%
Using the experimental values, $\dfrac{\alpha }{\alpha _{3}}=.0674$ and $%
\sin ^{2}\theta _{W}=0.231,$ \cite{PDG} we get%
\begin{equation}
\overset{\_}{b}=0.71  \label{b-bar}
\end{equation}%
which is value needed in the unification to get the right value for $\sin
^{2}\theta _{W}.$ Note that in the unification of the Standard Model without
supersymmetry, we have $\overset{\_}{b}=\dfrac{1}{2}$ which gives $\sin
^{2}\theta _{W}=0.204,$ and is significantly different from the experimental
value. The inclusion of Higgs scalars will have a very small effect and will
not change this result significantly. Or equivalently, if we use the
experimental value for $\sin ^{2}\theta _{W}$, coupling constants, when the
extrapolated to the unification scales, will not converge to a single
coupling constant. Comparing the Standard Model value with Eq (\ref{b-bar})
we see that we need to increase $\overset{\_}{b}$ in the Standard Model to
satisfy the experimental value.

Recall that various contributions to the $\beta -$function for the gauge
coupling constants are of the form,%
\begin{equation}
\beta _{i}=\dfrac{-g_{i}^{3}}{16\pi ^{2}}\left[ \dfrac{11}{3}t_{2}^{\left(
i\right) }\left( V\right) -\dfrac{2}{3}t_{2}^{\left( i\right) }\left(
F\right) -\dfrac{1}{3}t_{2}^{\left( i\right) }\left( S\right) \right]
\label{beta1}
\end{equation}%
Here $t_{2}^{\left( i\right) }\left( V\right) ,t_{2}^{\left( i\right)
}\left( F\right) $ and $t_{2}^{\left( i\right) }\left( S\right) $ are the
contributions to $\beta _{i}$ from gauge bosons, Weyl fermions, and complex
scalars respectively. For fermions in the representation with representation
matrices $T^{a}$, they are given by%
\begin{equation}
t_{2}\left( F\right) \delta ^{ab}=Tr\left\{ T^{a}\left( F\right) T^{b}\left(
F\right) \right\}
\end{equation}%
where the trace means summing over all members of the multiplets. Similarly,
for the gauge bosons and scalar contributions, $t_{2}\left( V\right) $ and $%
t_{2}\left( S\right) .$ For example, for the gauge bosons, $T^{a\prime }s$
are the matrices for the adjoint representations and for the gauge group $%
SU\left( n\right) ,$ $t_{2}^{\left( n\right) }\left( V\right) $ is given by,%
\begin{eqnarray}
t_{2}^{\left( n\right) }\left( V\right) &=&n,\qquad \text{for }n>2,\qquad
\label{Casmir} \\
t_{2}^{\left( n\right) }\left( V\right) &=&0,\qquad \text{for }n=1  \notag
\end{eqnarray}%
Note that $t_{2}^{\left( i\right) }\left( R\right) ,$ $R=V,$ $S,$ $F$ are
pure group theory factors and are independent of the spin of the particles.
For the low rank representations in the $SU\left( n\right) $ group they are
given by,

\begin{equation*}
\begin{tabular}{|c|c|}
\hline
& $t_{2}^{\left( n\right) }$ \\ \hline
$\text{fundamental rep}$ & $\dfrac{1}{2}$ \\ \hline
$\text{adjoint rep}$ & $n$ \\ \hline
$\text{2nd rank antisymmetric tensor}$ & $\dfrac{n-2}{2}$ \\ \hline
$\text{2nd rank antisymmetric tensor}$ & $\dfrac{n+2}{2}$ \\ \hline
\end{tabular}%
\end{equation*}%
Since $\dfrac{1}{16\pi ^{2}}$ is common to all $\beta -$functions and will
be cancelled out in the ratio in $\overset{\_}{b}$ in Eq (\ref{b-bar1}) we
will neglect it in writing out the coefficient $b_{n},$%
\begin{equation*}
b_{n}=\left[ \dfrac{11}{3}t_{2}^{\left( n\right) }\left( V\right) -\dfrac{2}{%
3}t_{2}^{\left( n\right) }\left( F\right) -\dfrac{1}{3}t_{2}^{\left(
n\right) }\left( S\right) \right]
\end{equation*}

For convenience, we can use the parameter $\overset{\_}{b}$ given in Eq (\ref%
{b-bar}) to discuss the contribution to the Weinberg angle $\theta _{W}$
from models with new particles.%
\begin{equation}
\overset{\_}{b}=\dfrac{b_{3}{}_{2}}{b_{21}}  \label{b-bar2}
\end{equation}%
where%
\begin{equation}
b_{32}\equiv b_{3}-b_{2},\qquad b_{21}\equiv b_{2}-b_{1}  \label{b-bar3}
\end{equation}%
We will separate the contributions into gauge bosons, fermions, and scalars,%
\begin{equation}
b_{32}=b_{32}\left( V\right) +b_{32}\left( F\right) +b_{32}\left( S\right)
\end{equation}%
\begin{equation}
b_{21}=b_{21}\left( V\right) +b_{21}\left( F\right) +b_{21}\left( S\right)
\end{equation}%
Note that for a complete multiplet of $SU\left( 5\right) ,$ their
contributions to $b_{32}$ and $b_{21}$ cancel out in the combination in Eq (%
\ref{b-bar2}). This can be seen as follows. Let $T^{\left( a\right) }\left(
R\right) $ be the matrices for some representation $R$ of \ $SU\left(
5\right) $ group. Then the coefficient $t_{2}\left( R\right) $ of the second
Casmir operator given by%
\begin{equation}
t_{2}\left( R\right) \delta ^{ab}=tr\left\{ T^{\left( a\right) }\left(
R\right) T^{\left( b\right) }\left( R\right) \right\}
\end{equation}%
will be the same when we take $T^{\left( a\right) }\left( R\right) \in $ $%
SU\left( 3\right) ,$ $SU\left( 2\right) $ or $U\left( 1\right) $, 
\begin{equation}
t_{2}^{\left( 3\right) }\left( R\right) =t_{2}^{\left( 2\right) }\left(
R\right) =t_{2}^{\left( 1\right) }\left( R\right)
\end{equation}%
Thus their contributions will cancel out in the combination $b_{3}-b_{2}$ or 
$b_{2}-b_{1}.$ In other words, a complete multiplet of $SU\left( 5\right) ,$
e.g. $\mathbf{5}$ or $\mathbf{10,}$ will not effect the prediction of the
Weinberg angle, even though they will effect the evolution of each
individual coupling constant. Here we assume that all members of the $%
SU\left( 5\right) $ multiplets survive to low energies. This is the case for
the fermion contributions in the Standard Model. Thus for the studies of the
Weinberg angle, we need to consider only the incomplete $SU\left( 5\right) $
multiplets, i.e. those $SU\left( 3\right) \times SU\left( 2\right) \times
U\left( 1\right) $ multiplets which do not combine into a complete $SU\left(
5\right) $ multiplets. The reason that we can have the incomplete multiplets
in the unification theory has to do with the decoupling of superheavy
particles. For example, in the standard $SU\left( 5\right) $ unification,
only gauge bosons in subgroups $SU\left( 3\right) \times SU\left( 2\right)
\times U\left( 1\right) $ survive to low energies, while the other gauge
bosons, e.g. leptoquarks, are superheavy and decouple. Similarly, only Higgs
bosons in $SU\left( 2\right) $ doublet will contribute to the $\beta -$%
function.

\section{Contribution to $\protect\theta _{W}$ from Various Multiplets}

The contributions of $SU\left( 3\right) \times SU\left( 2\right) \times
U\left( 1\right) $ gauge bosons are given by, using Eq (\ref{Casmir}) 
\begin{equation}
b_{3}\left( V\right) =11,\qquad b_{2}\left( V\right) =\dfrac{22}{3},\qquad
b_{1}\left( V\right) =0
\end{equation}%
and they give%
\begin{equation}
b_{32}\left( V\right) =\dfrac{11}{3},\qquad b_{21}\left( V\right) =\dfrac{22%
}{3}
\end{equation}%
The contribution from the Higgs scalars are%
\begin{equation}
b_{32}\left( S\right) =\dfrac{1}{6},\qquad b_{21}\left( S\right) =-\dfrac{1}{%
15}
\end{equation}%
Since these contributions are very small, they are not included in the usual
analysis.

We now consider the effects of various low rank multiplets to the Weinberg
angle, or the parameter $\overset{\_}{b}$ in Eq (\ref{b-bar2}). The results
for the scalars and fermions are listed separately in the following tables:

(a) Scalars:

\begin{equation*}
\begin{tabular}{|c|c|c|c|c|c|c|}
\hline
$SU\left( 3\right) $ & $SU\left( 2\right) $ & $b_{1}\left( s\right) $ & $%
b_{2}\left( s\right) $ & $b_{3}\left( s\right) $ & $b_{32}\left( S\right) $
& $b_{21}\left( S\right) $ \\ \hline
color singlet & doublet \ $\left( 
\begin{array}{c}
\phi ^{0} \\ 
\phi ^{-}%
\end{array}%
\right) ,$ & $-\dfrac{1}{10}$ & $-\dfrac{1}{6}$ & $0$ & $\dfrac{1}{6}$ & $-%
\dfrac{1}{15}$ \\ \hline
$^{\prime \prime }$ & triplet \ \ $\left( 
\begin{array}{c}
\phi ^{+} \\ 
\phi ^{0} \\ 
\phi ^{-}%
\end{array}%
\right) ,$ & $0$ & $-\dfrac{2}{3}$ & $0$ & $\dfrac{2}{3}$ & $-\dfrac{2}{3}$
\\ \hline
$^{\prime \prime }$ & triplet \ \ $\left( 
\begin{array}{c}
\phi ^{++} \\ 
\phi ^{+} \\ 
\phi ^{0}%
\end{array}%
\right) ,$ & $-\dfrac{3}{5}$ & $-\dfrac{2}{3}$ & $0$ & $\dfrac{2}{3}$ & $-%
\dfrac{1}{15}$ \\ \hline
$^{\prime \prime }$ & Singlet \ $\phi ^{+}$, \ \ \ \ \ \ \ \ \ \  & $-\dfrac{%
1}{5}$ & $0$ & $0$ & $0$ & $\dfrac{1}{5}$ \\ \hline
$^{\prime \prime }$ & Singlet \ $\phi ^{++}$, \ \ \ \ \ \ \ \  & $-\dfrac{4}{%
5}$ & $0$ & $0$ & $0$ & $\dfrac{4}{5}$ \\ \hline
$^{\prime \prime }$ & Singlet \ $\phi ^{0}$, \ \ \ \ \ \ \ \ \  & $0$ & $0$
& $0$ & $0$ & $0$ \\ \hline
color triplet & doublet \ $\left( 
\begin{array}{c}
\NEG{u} \\ 
\NEG{d}%
\end{array}%
\right) _{L}$ \  & $-\dfrac{1}{30}$ & $-\dfrac{1}{2}$ & $-\dfrac{1}{3}$ & $%
\dfrac{1}{6}$ & $-\dfrac{7}{15}$ \\ \hline
$^{\prime \prime }$ & singlet \ \ \ $\NEG{u}$ & $-\dfrac{4}{15}$ & $0$ & $-%
\dfrac{1}{6}$ & $-\dfrac{1}{6}$ & $\dfrac{4}{15}$ \\ \hline
$^{\prime \prime }$ & Singlet \ \ \ $\NEG{d}$ & $-\dfrac{1}{15}$ & $0$ & $-%
\dfrac{1}{6}$ & $-\dfrac{1}{6}$ & $\dfrac{1}{15}$ \\ \hline
$^{\prime \prime }$ & 1 generation of squarks & $-\dfrac{11}{30}$ & $-\dfrac{%
1}{2}$ & $-\dfrac{2}{3}$ & $-\dfrac{1}{6}$ & $-\dfrac{2}{15}$ \\ \hline
\end{tabular}%
\end{equation*}

(b) Fermions:%
\begin{equation*}
\begin{tabular}{|c|c|c|c|c|c|c|}
\hline
$SU\left( 3\right) $ & $SU\left( 2\right) $ & $b_{1}\left( f\right) $ & $%
b_{2}\left( f\right) $ & $b_{3}\left( f\right) $ & $b_{32}\left( f\right) $
& $b_{21}\left( f\right) $ \\ \hline
color singlet & doublet \ $\left( 
\begin{array}{c}
\nu \\ 
l^{-}%
\end{array}%
\right) $ & $-\dfrac{1}{5}$ & $-\dfrac{1}{3}$ & $0$ & $\dfrac{1}{3}$ & $-%
\dfrac{2}{15}$ \\ \hline
$^{\prime \prime }$ & triplet \ \ $\left( 
\begin{array}{c}
l^{+} \\ 
l^{0} \\ 
l^{-}%
\end{array}%
\right) $ & $0$ & $-\dfrac{4}{3}$ & $0$ & $\dfrac{4}{3}$ & $-\dfrac{4}{3}$
\\ \hline
$^{\prime \prime }$ & triplet \ \ $\left( 
\begin{array}{c}
l^{++} \\ 
l^{+} \\ 
l^{0}%
\end{array}%
\right) $ & $-\dfrac{6}{5}$ & $-\dfrac{4}{3}$ & $0$ & $\dfrac{4}{3}$ & $-%
\dfrac{2}{15}$ \\ \hline
$^{\prime \prime }$ & singlet \ \ \ \ \ \ \ \ \ \ $l^{+}$ & $-\dfrac{2}{5}$
& $0$ & $0$ & $0$ & $\dfrac{2}{5}$ \\ \hline
$^{\prime \prime }$ & singlet \ \ \ \ \ \ \ \ \ \ $l^{0}$ & $0$ & $0$ & $0$
& $0$ & $0$ \\ \hline
color triplet & doublet \ $\left( 
\begin{array}{c}
u \\ 
d%
\end{array}%
\right) $ & $-\dfrac{1}{15}$ & $-1$ & $-\dfrac{2}{3}$ & $\dfrac{1}{3}$ & $-%
\dfrac{14}{15}$ \\ \hline
$^{\prime \prime }$ & singlet \ \ \ $\ \ \ \ \ u_{R}$ & $-\dfrac{8}{15}$ & $%
0 $ & $-\dfrac{1}{3}$ & $-\dfrac{1}{3}$ & $\dfrac{8}{15}$ \\ \hline
$^{\prime \prime }$ & singlet \ \ \ $\ \ \ \ \ d_{R}$ & $-\dfrac{2}{15}$ & $%
0 $ & $-\dfrac{1}{3}$ & $-\dfrac{1}{3}$ & $\dfrac{2}{15}$ \\ \hline
$^{\prime \prime }$ & 1 generation of quarks & $-\dfrac{11}{15}$ & $-1$ & $-%
\dfrac{4}{3}$ & $-\dfrac{1}{3}$ & $-\dfrac{4}{15}$ \\ \hline
color octet & glunios & $0$ & $0$ & $-2$ & $-2$ & $0$ \\ \hline
\end{tabular}%
\end{equation*}%
Remarks :As it is evident in Eq (\ref{beta1}) with same $SU\left( 3\right)
\times SU\left( 2\right) \times U\left( 1\right) $ quantum numbers the
contribution from the scalars are $\dfrac{1}{2}$ of the fermions
contributions.

\bigskip

\section{\textbf{Supersymmetric extension of standard model and
generalization}}

As we have discussed before, we need to consider only the incomplete
multiplets. The particles needed to be included are given below,

\begin{enumerate}
\item Scalars:%
\begin{equation*}
2\text{ Higgs doublets\qquad }H=\left( 
\begin{array}{c}
H^{+} \\ 
H^{0}%
\end{array}%
\right) \qquad b_{32}\left( H\right) =\dfrac{1}{3},\text{ \ \ \ }%
b_{21}\left( H\right) =-\dfrac{2}{15}
\end{equation*}

\item Fermions%
\begin{equation*}
\begin{array}{cccc}
\text{glunios\ } & \tilde{g}^{a},a=1,\ldots 8\text{ \ \ \ \ \ \ \ } & 
b_{32}\left( \tilde{g}^{a}\right) =-2\qquad & b_{21}\left( \tilde{g}%
^{a}\right) =0 \\ 
\text{Winos} & \tilde{W}=\left( 
\begin{array}{c}
\tilde{W}^{+} \\ 
\tilde{W}^{0} \\ 
\tilde{W}^{-}%
\end{array}%
\right) & b_{32}\left( \tilde{W}\right) =\dfrac{4}{3} & b_{21}\left( \tilde{W%
}\right) =-\dfrac{4}{3} \\ 
\text{binos} & \tilde{b} & b_{32}\left( \tilde{b}\right) =0 & b_{21}\left( 
\tilde{b}\right) =0 \\ 
2\text{ Higginos} & \tilde{h}=\left( 
\begin{array}{c}
\tilde{h}^{0} \\ 
\tilde{h}^{-}%
\end{array}%
\right) & b_{32}\left( \tilde{h}\right) =\dfrac{2}{3} & b_{21}\left( \tilde{h%
}\right) =-\dfrac{4}{15}%
\end{array}%
\end{equation*}
\end{enumerate}

Thus the total contributions are 
\begin{equation}
b_{32}=\dfrac{1}{3},\qquad b_{21}=-\dfrac{26}{15}  \label{susy-b}
\end{equation}%
and using Eq (\ref{b-bar2}), we get for $\overset{\_}{b},$%
\begin{equation}
\overset{\_}{b}=\dfrac{\dfrac{11}{3}+\dfrac{1}{3}}{\dfrac{22}{3}-\dfrac{26}{%
15}}=.715
\end{equation}%
This agrees with the experimental measurement very well and gives a strong
support for the supersymmetric extension of Standard Model. Note that the
result $b_{32}=\dfrac{1}{3}$ comes from a cancellation between large glunios
contribution and that of winos and Higginos. Thus the mechanism to produce
the right value for the Weinberg angle in the supersymmetric extension of \
Standard Model is quite complicate. This makes the SSM somewhat unique in
the sense the extra particles which give all these cancellations are not
arbitrary but are required by the principle of supersymmetry.

\qquad We will now discuss the other examples which can give satisfactory
answer for the Weinberg angle. Since the supersymmetric extension of
Standard Model has been so successful in the prediction of Weinberg angle,
we will now give the possibilities of other multiplets which give the same
contribution. We will use a simpler notation for the multiplet structure, $%
\left( \mathbf{a,b}\right) _{Y}^{f}$ , denote a fermion multiplet transforms
as representations $\mathbf{a,b}$ under $SU\left( 3\right) \times SU\left(
2\right) $ with hypercharge $Y.$ Similarly, $\left( \mathbf{a,b}\right)
_{Y}^{S}$ for the scalar multiplets. To present the result, we group
together the multiplets with same contributions for the coefficients
functions, $b_{3},b_{2},$ and $b_{1}.$

\begin{enumerate}
\item $b_{32}=-2,$ $b_{21}=0$ (glunio-like states)%
\begin{equation*}
\begin{tabular}{|c|c|c|c|}
\hline
$\left( \mathbf{8,1}\right) _{0}^{f}$ & $\left( \mathbf{6,1}\right)
_{0}^{f}\oplus \left( \overset{\_}{\mathbf{3}}\mathbf{,1}\right) _{0}^{f}$ & 
$\left( \mathbf{3,1}\right) _{0}^{f}\times 6$ & $\left( \mathbf{8,1}\right)
_{0}^{s}\times 2$ \\ \hline
$\left( \mathbf{8,1}\right) _{0}^{s}\oplus \left( \mathbf{6,1}\right)
_{0}^{s}\left( \mathbf{3,1}\right) _{0}^{s}$ & $\left( \mathbf{6,1}\right)
_{0}^{s}\times 2\oplus \left( \overset{\_}{\mathbf{3}}\mathbf{,1}\right)
_{0}^{f}$ & $\left( \mathbf{6,1}\right) _{0}^{s}\oplus \left( \overset{\_}{%
\mathbf{3}}\mathbf{,1}\right) _{0}^{f}\times 2$ &  \\ \hline
\end{tabular}%
\end{equation*}

\item $b_{32}=\dfrac{4}{3},$ $b_{21}=-\dfrac{4}{3}$ (Wino-like states)%
\begin{equation*}
\begin{tabular}{|c|c|c|}
\hline
$\left( \mathbf{1,3}\right) _{0}^{f}$ & $\left( \mathbf{1,3}\right)
_{0}^{s}\times 2$ & $\left( \mathbf{1,2}\right) _{0}^{s}\times 8$ \\ \hline
\end{tabular}%
\end{equation*}

\item $b_{32}=\dfrac{2}{3},$ $b_{21}=-\dfrac{4}{15}$ (Higgio-like states)%
\begin{equation*}
\begin{tabular}{|c|c|c|}
\hline
$\left( \mathbf{1,2}\right) _{-1}^{f}\times 2$ & $\left( \mathbf{1,2}\right)
_{-1}^{s}\times 4$ & $\left( \mathbf{1,3}\right) _{0}^{s}\oplus \left( 
\mathbf{1,1}\right) _{2}^{f}$ \\ \hline
$\left( \mathbf{1,2}\right) _{-1}^{f}\oplus \left( \mathbf{1,2}\right)
_{-1}^{s}\times 2$ & $\left( \mathbf{1,3}\right) _{0}^{s}\oplus \left( 
\mathbf{1,1}\right) _{2}^{s}\times 2$ &  \\ \hline
\end{tabular}%
\end{equation*}

\item $b_{32}=\dfrac{1}{3},$ $b_{21}=-\dfrac{2}{15}$ (Higgs-like states)%
\begin{equation*}
\begin{tabular}{|c|}
\hline
$\left( \mathbf{1,2}\right) _{-1}^{s}\times 2$ \\ \hline
\end{tabular}%
\end{equation*}%
Note that this last group includes the Standard Model Higgs contribution.
\end{enumerate}

Clearly, if take one set of multiplet from each group, the total
contribution will give the same Weinberg angle as the supersymmetry Standard
Model. The list given above is constructed by mimicking those multiplets of
the supersymmetry Standard Model and are by no means the only possibilities.
We just want to demonstrate the existence of other examples. None of these
possibilities, except for the supersymmetric one, are realized in any
realist phenomenological model. For the case of new fermions multiplets, one
might worry about the anomaly cancellation. However, as we have discussed
before, only the incomplete multiplets are relevant for the Weinberg angle
and it is possible to cancel the anomalies by fermion multiplets which
decouple at very high energies.

\end{document}